\documentclass[aps,prd,a4paper,onecolumn,amsmath,showpacs,superscriptaddress,nofootinbib,preprintnumbers,notitlepage]{revtex4-1}
 
\usepackage{verbatim}
\usepackage[T1]{fontenc}
\usepackage[utf8]{inputenc}
\usepackage[american]{babel}
\usepackage{epsfig}
\usepackage{graphicx,subcaption,caption}
\usepackage{pgfplots}
\usepackage{booktabs}
\usepackage{multirow}
\usepackage{dcolumn}
\usepackage{amsmath}
\usepackage{mathtools}
\usepackage{amsfonts}
\usepackage{amssymb}
\usepackage{epstopdf}
\usepackage{bm}
\usepackage{siunitx}
\usepackage{braket}
\usepackage{enumitem}
\usepackage{soul}
\usepackage{color}
\usepackage{transparent}
\usepackage{pifont}

\definecolor{navyblue}{rgb}{0.0, 0.0, 0.5}
\definecolor{royalblue}{rgb}{0.25, 0.41, 0.88}
\definecolor{cadmiumgreen}{rgb}{0.0, 0.42, 0.24}
\definecolor{blue-violet}{rgb}{0.54, 0.17, 0.89}
\definecolor{darkviolet}{rgb}{0.58, 0.0, 0.83}
\definecolor{orange(colorwheel)}{rgb}{1.0, 0.5, 0.0}

\usepackage{hyperref}
\hypersetup{
    colorlinks=true, 
    linkcolor=royalblue, 
    citecolor=magenta}

\newcommand\be{\begin{equation}}
\newcommand\ee{\end{equation}}

\newcommand\bea{\begin{eqnarray}}
\newcommand\eea{\end{eqnarray}}

\usepackage{booktabs}
\usepackage{multirow}
\usepackage{dcolumn}
\usepackage{colortbl}

\definecolor{magenta(process)}{rgb}{1.0, 0.0, 0.56}

\definecolor{darkspringgreen}{rgb}{0.09, 0.45, 0.27}

\definecolor{royalblue(web)}{rgb}{0.25, 0.41, 0.88}

\begin{document}

\title{The generalized $sl(2, R)$ and $su(1, 1)$ in non-minimal constant-roll inflation}

\author{Mehdi Shokri}
\email{mehdishokriphysics@gmail.com}
\affiliation{Department of Physics, Campus of Bijar, University of Kurdistan, Bijar, Iran}

\author{Jafar Sadeghi}
\email{pouriya@ipm.ir}
\affiliation{Department of Physics, University of Mazandaran, P. O. Box 47416-95447, Babolsar, Iran}

\author{Mohammad Reza Setare}
\email{rezakord@ipm.ir}
\affiliation{Department of Physics, Campus of Bijar, University of Kurdistan, Bijar, Iran}

\preprint{}
\begin{abstract}
In the present work, we consider the non-minimal coupling inflationary model in the context of the constant-roll idea which is investigated by the first-order formalism. We attempt to find the hidden symmetries behind the model by the Lie symmetry method. We supply this aim by using the symmetry features of the Heun function instead of Killing vector approach. We show that the hidden symmetries of the non-minimal constant-roll inflation in the cases of power-law and exponential couplings are characterized as a generalized form of $sl(2, R)$ and $su(1, 1)$ algebra, respectively. 
\end{abstract}

\maketitle
\section{Introduction}
Cosmic inflation has been recognized as the most well-defined theory to describe the early universe phenomena by generating the primordial scalar and tensor perturbations responsible for structure formation of the universe and the primordial gravitational waves, respectively \cite{Guth,Starobinsky:1980te,Linde:1981my,Albrecht:1982wi}. As the simplest inflationary model, we deal with a single scalar field (inflaton) rolling down slowly from the top of potential to the minimum point due to the slow-roll approximation. Then, inflation ends when inflaton decays at the final phase of inflation throughout a reheating process \cite{Kofman2}. Despite the successes of the single field model, it suffers from the lack of non-Gaussinaity in its spectrum due to uncorrelated modes of the spectrum \cite{Chen}. This can be problematic when our future observations present the non-Gaussianity in the perturbations spectrum. Going beyond the slow-roll approximation is proposed to create some non-Gaussianities in the perturbations spectrum of the single field models in order to remove the likely shortcomings. Hence, a new approach to the inflationary paradigm is introduced by considering a scalar field with a constant rate of rolling during the inflationary era as $\ddot{\varphi}=\beta H\dot{\varphi}$
where $\beta=-(3+\alpha)$ and $\alpha$ is a non-zero parameter \cite{martin2,Motohashi1,Motohashi2}. Variations from the slow-roll approximation also can be discovered in ultra slow-roll inflation where we deal with a non-negligible $\ddot{\varphi}$ in the Klein-Gordon equation as  $\ddot{\varphi}=3H\dot{\varphi}$. This class of inflationary models is situated in the non-attractor phase of inflation with a scale-invariant spectrum. Also, the non-Gaussinaity consistency relation of single field models is violated in the presence of ultra condition through Super-Hubble evolution of the scalar perturbation \cite{Namjoo}. As another class of models proposed in the beyond of the slow-roll approximation, we can introduce the fast-roll models in which a fast-rolling stage is considered at the start of inflation and will be attached to the standard slow-roll only after a few e-folds \cite{Contaldi}. Recently, Constant-roll inflation has been engaged for a wide range of inflationary models since it has opened a new window to understand cosmic inflation by a different initial condition \cite{Odintsov,Nojiri,Motohashi9,Morse,Ghersi,Lin,Micu,Kamali}.

On the other hand, we are enthusiastic to search the existing symmetries of the inflationary models introduced in the context of constant-roll inflation. For this reason, we work with the concept of the Lie symmetries as a main mathematical tool for studying the nonlinear differential equations \cite{Ovsiannikov,Ibragimov,Olver,Crampin,Leach}. In fact, Lie's theory helps us to find the exact and analytic solutions of many gravitational field equations and also the cosmological situations \cite{Ritis,Christodoulakis,Capo,Dimakis}. An important class of the Lie symmetries commonly used in gravitational theories are the space-time collineations. In that case, they are introduced mostly by Killing vectors as the generators of continuous transformations where geometric objects of the gravitational theory are transformed under a specific rule or remain invariant. Generally, finding the hidden symmetries of a model which are obtained by the corresponding algebra, helps us to be more familiar with the physical features of that model.

In the present work, we attempt to trace the hidden symmetries of the non-minimal constant-roll inflationary model by the Lie symmetry approach. In \cite{sadeghi}, we have investigated a single field inflationary model equipped with a non-minimal coupling term between the Ricci scalar $R$ and inflaton $\varphi$ in the context of constant-roll inflation by the first-order formalism. The method is very useful to study cosmological situations using the obtained potential of the scalar field \cite{Bazeia1,Bazeia2,Bazeia3,Bazeia4,Bazeia5,Bazeia6,Bazeia7}. This can be performed by offering two main functions $W=W(\varphi)$ and $Z=Z(\varphi)$ which leads to reduce the equations of motion to the first-order differential equations.  Since the obtained differential equations of the model take the Heun function form, we use the symmetry features of the Heun function instead of the Killing vector approach. The confluent and double confluent Heun equations with some applications are discussed in \cite{Decarreau1,Decarreau2,Figueiredo1,Figueiredo2,Figueiredo3,Kimura,Leaver}. The Heun functions also are known as the generalized wave equation \cite{Ronveaux,Wilson1,Wilson2}. As we can see in \cite{Teitelboim,Maldacena}, the general form of the Heun equation takes the following form
\begin{equation}
y''(x)+(\frac{\eta}{x}+\frac{\delta}{x-1}+\sum^{k}_{i=1}\frac{\epsilon_{i}}{x-a_{i}})y'(x)+
\frac{\mu\nu x^{k}+\sum^{k}_{i=1}p_{i}x^{k-1}}{x(x-1)\prod^{k}_{i=1}(x-a_{i})}y(x)=0
\label{1}
\end{equation}
where two indices at each singularity $a_{i}$ are $(0,1-\epsilon_{i})$, $(0,1-\eta)$, $(0,1-\delta)$ and $(\mu,\nu)$ at $x=0$, $x=1$ and $x=\infty$, respectively. Now, by taking into account the Fuchsian relation, we have 
\begin{equation}
\mu+\nu+1=\eta+\delta+\sum^{k}_{i=1}\epsilon_{i}.
\label{2}
\end{equation}
In order to follow the main purpose of the paper, we write the equation (\ref{1}) in terms of $f_{i}(x)(i=1,...,k)$ functions and then by comparing with respect to the obtained equations of the model, we calculate the special operators $P^{+}$, $P^{-}$ and $P^{0}$. Finally, we show such operators are satisfied by the commutation relations of generators as $J^{+}$, $J^{-}$, and $J^{0}$. In fact, the commutation relations between the three generators allow us to understand the special symmetry algebra behind the model. The above discussion motivates us to arrange the paper as follows. In section II, we present a brief explanation about the Heun function and also the method used to obtain the corresponding operators. In section III, we apply the method introduced in the previous section for the obtained results of non-minimal constant-roll inflation in two different cases of coupling i.e. power-law and exponential couplings. Finally, we conclude the analysis in section IV.
\section{A review of the Heun equation}
Because of our purpose, we begin with the general form of the Heun equation (\ref{1}) in the case of $k=4$ \cite{sadeghi} as follows 
\begin{equation}
y''(x)+(\frac{\eta}{x}+\frac{\delta}{x-1}+\frac{\epsilon_{1}}{x-a}+\frac{\epsilon_{2}}{x-b}+
\frac{\epsilon_{3}}{x-c}+\frac{\epsilon_{4}}{x-d})y'(x)+
\frac{\mu\nu x^{4}+p_{1}+p_{2}x+p_{3}x^{2}+p_{4}x^{3}}{x(x-1)(x-a)(x-b)(x-c)(x-d)}y(x)=0.
\label{3}
\end{equation}
Also, we can write the above differential equation in a general form in terms of $f_{1}(x) $, $f_{2}(x)$ and $f_{3}(x)$ as
\begin{equation}
f_{1}(x)y''(x)+f_{2}(x)y'(x)+f_{3}(x)y(x)=0
\label{4}
\end{equation}
where $f_{1}(x) $, $ f_{2}(x)$ and $ f_{3}(x)$ are polynomial functions as
\begin{eqnarray}
&\!&\!f_{1}(x)=a_{0}x^{6}+a_{1}x^{5}+a_{2}x^{4}+a_{3}x^{3}+a_{4}x^{2}+a_{5}y+a_{6}\nonumber\\&\!&\!
f_{2}(x)=a_{7}x^{5}+a_{8}x^{4}+a_{9}x^{3}+a_{10}x^{2}+a_{11}x+a_{12}
\nonumber\\&\!&\!
f_{3}(x)=a_{13}x^{4}+a_{14}x^{3}+a_{15}x^{2}+a_{16}x+a_{17}.
\label{5}
\end{eqnarray}
By comparing the equation (\ref{4}) with the equation (\ref{3}), one can obtain the coefficients $a_{i}$ (with $i=0,...17$) in terms of the Heun equation parameters. Now, we build the algebra operators $P^{+}$, $P^{-}$, $P^{0}$ and $F(x\frac{d}{dx})$ as,
\begin{equation}
P^{+}=a_{0}x^{6}\frac{d^{2}}{dx^{2}}+a_{7}x^{5}\frac{d}{dx}+a_{13}x^{4},\quad\quad
P^{-}=a_{2}x^{4}\frac{d^{2}}{dx^{2}}+a_{9}x^3\frac{d}{dx}+a_{15}x^2,\quad\quad F(x\frac{d}{dx})=a_{1}x^{5}\frac{d^{2}}{dx^{2}}+a_{8}x^{4}\frac{d}{dx}+a_{14}x^{3}.
\label{6}
\end{equation}
Notice that the form of $P^{0}$ will be determined through the corresponding symmetry algebra of the model. For example, for having any form of $sl(2)$ algebra, we consider $P^{0}=\mu x\frac{d}{dx}+\nu$ so that the parameters $\mu$ and $\nu$ take the special values satisfying the following commutation relations
\begin{equation}
[P^{0},P^{+}]= P^{0},\quad\quad\quad [P^{0},P^{-}]=-P^{-}
,\quad\quad\quad [P^{+},P^{-}]=FP^{0}.
\label{7}
\end{equation}
As we know the Heun function gives us some symmetries of the model, now it is time to use the method for the non-minimal coupling constant-roll inflationary model discussed in \cite{sadeghi}.
\section{The non-minimal constant-roll inflation}
As the target model, we examine the non-minimal coupling inflationary model expressed in the context of the constant-roll approach. According to the non-minimal coupling idea, a non-minimal interaction between two fields the Ricci scalar $R$ and scalar field $\varphi$ is unavoidable in different cosmological scenarios which deal with a scalar field. Hence, the standard inflationary action modified by a non-minimal term can be introduced as follows
\begin{equation}
S_{J}=\int{d^{4}x\sqrt{-g}\bigg(\frac{f(\varphi)R}{2}-\frac{1}{2}g^{\mu\nu}\partial_{\mu}\varphi\partial_{\nu}\varphi-V(\varphi)\bigg)}
\label{8}
\end{equation}
where $g$ is the determinant of the metric $g_{\mu\nu}$ and $R=g^{\mu\nu}R_{\mu\nu}$ is the Ricci scalar. Also, here we suppose $\kappa^{2}\equiv8\pi G=1$. By using the energy-momentum tensor of a perfect fluid and Friedmann-Robertson-Walker (FRW) metric $ds^{2}=-dt^{2}+a(t)^{2}(dx^{2}+dy^{2}+dz^{2})$ for a spatially flat universe which is homogeneous and isotropic, the dynamical equations are given by
\begin{equation}
3fH^{2}=\frac{\dot{\varphi}^{2}}{2}+V-3Hf'\dot{\varphi},
\label{9}
\end{equation}
\begin{equation}
2f\dot{H}=-(1+f'')\dot{\varphi}^{2}-f'(\ddot{\varphi}-H\dot{\varphi})
\label{10}
\end{equation}
where $H\equiv\frac{\dot{a}}{a}$ is the Hubble parameter, and a dot and a prime represent the derivative with respect to cosmic time and $\varphi$, respectively. Moreover, by varying the action (\ref{8}) with respect to the $\varphi$, the Klein-Gordon equation as the equation of motion of the scalar field is found by,
\begin{equation}
\ddot{\varphi}+3H\dot{\varphi}+\frac{d V}{d\varphi}-3f'(2H^{2}+\dot{H})=0.
\label{11}
\end{equation}
Obviously, the above expressions will be reduced to the standard case when $f=1$.

In order to investigate the non-minimal constant-roll inflation, we engage the first-order formalism by introducing two parameters $W(\varphi)$ and $Z(\varphi)$ which reduces the equations to the first-order differential equations. The corresponding method can be applied for a wide range of cosmological situations since it directly relates the function $W$ with Hubble’s parameter $H$. As we can see in \cite{sadeghi}, the simplest form of first-order differential equation of the scalar field 
\begin{equation}
H=W(\varphi)  
\label{12}
\end{equation}
leads to some linear differential equations with trivial inflationary solutions. However, the assumption (\ref{12}) sometimes is not sufficient to solve the problems of the model particularly when we deal with a standard formalism modified by some quantum corrections. Hence, we require to consider a new factor $Z(\varphi)$ in addition to $W(\varphi)$ so that the relation (\ref{12}) is modified as
\begin{equation}
H=W(\varphi)+\mathcal{T} Z(\varphi)
\label{13}
\end{equation}
where $\mathcal{T}$ is a constant determined by the coupling form of the model.
\subsection{Power-law coupling}
First, we consider the power-law coupling as the simplest form of coupling widely-used in many cosmological literature as
\begin{equation}
f(\varphi)=1-\xi\varphi^{2}
\label{14}
\end{equation}
where $\xi$ is coupling constant and its value profoundly affects the viability of inflationary models. By using the equation (\ref{13}) in which $\mathcal{T}$ is defined as $\mathcal{T}=\gamma\xi$ and also the constant-roll condition $\ddot{\varphi}=\beta H\dot{\varphi}$, we obtain the constraint equation as a Heun function which is expressed by \cite{sadeghi}
\begin{equation}
\gamma\xi(1-\xi\varphi^{2})W''+(-\xi\varphi^{2}+\gamma\xi^{2}\varphi(-\beta-2+6\xi)+1)W'+\varphi\xi(-\beta-2+6\xi)W=0. 
\label{15}
\end{equation}
Let $\xi\varphi^{2}=x^{2}$ and $W=x\mathcal{W}$, then the above equation can be written by
\begin{equation}
x(1-x^{2})\mathcal{W}''+\bigg(-x^{3}+\gamma\xi^{3/2}x^{2}(-\beta-4+6\xi)+x+2\gamma\xi^{3/2}\bigg)\frac{\mathcal{W}'}{\gamma\xi^{3/2}}+\bigg(x^{2}(-\beta-3+6\xi)+\gamma\xi^{3/2}x(-\beta-2+6\xi)+1\bigg)\frac{\mathcal{W}}{\gamma\xi^{3/2}}=0.
\label{16}
\end{equation}
By comparing the equation  (\ref{16}) with the form of the Heun equation considered in previous section (\ref{3}), one can find $\epsilon_{3}=\epsilon_{4}=p_{3}=p_{4}=0$. Hence, the polynomial functions $f_{1}(x) $, $ f_{2}(x)$ and $ f_{3}(x)$ introduced in the equation (\ref{5}) take the following form as
\begin{eqnarray}
&\!&\!f_{1}(x)=a_{0}x^{4}+a_{1}x^{3}+a_{2}x^{2}+a_{3}x+a_{4}\nonumber\\&\!&\!  
f_{2}(x)=a_{5}x^{3}+a_{6}x^{2}+a_{7}x+a_{8}\nonumber\\&\!&\!  
f_{3}(x)=a_{9}x^{2}+a_{10}x+a_{11}
\label{17}
\end{eqnarray}
where the coefficients $a_{i}$ (with $i=0,...,11$) are given by the equation (\ref{16}). In order to find the symmetry behind the model, we write the operators $P^{+}$, $P^{-}$ and $P^{0}$ as
\begin{equation}
P^{+}=-\frac{1}{\gamma\xi^{3/2}}x^{3}\frac{d}{dx}+\frac{(-\beta-3+6\xi)}{\gamma\xi^{3/2}}x^{2},\quad\quad\quad P^{-}=\frac{1}{\gamma\xi^{3/2}}x\frac{d}{dx}+\frac{1}{\gamma\xi^{3/2}},\quad\quad\quad P^{0}=\mu x^{2}\frac{d}{dx}+\nu x 
\label{18}
\end{equation} 
and then by introducing $\mu=1/\gamma^{2}\xi^{3}=1$ and $\nu=(\beta+3-6\xi)/{\gamma^{2}\xi^{3}}=1$, the commutation relations are obtained 
\begin{equation}
[P^{0},P^{-}]=-xP^{-},\quad\quad\quad [P^{0},P^{+}]=xP^{+},\quad\quad\quad [P^{+},P^{-}]=2xP^{0}.
\label{19}
\end{equation}
The above information tell us that the behaviour of the operators $P^{+}$, $P^{-}$ and $P^{0}$ is the same with the operators $J^{+}$, $J^{-}$ and $J^{0}$ of the $sl(2, R)$ algebra with some modifications. Hence, we find that the non-minimal constant-roll inflationary model in the case of power-law coupling obeys from a generalized form of $sl(2, R)$ algebra. Also, the quadratic Casimir operator $C$ which commutes with all existing operators of the model can be obtained by $C=(P^{0})^{2}+\frac{1}{2}(P^{+}P^{-}+P^{-}P^{+})$.
\subsection{Exponential coupling}
Now, we are going to account for another form of coupling studied in \cite{sadeghi}
\begin{equation}
f(\varphi)=e^{k\varphi} 
\label{20}
\end{equation} 
where $k$ is a constant. By using the first-order equation (\ref{13}) with $\mathcal{T}=\gamma k$ and also the constant-roll condition $\ddot{\varphi}=\beta H\dot{\varphi}$, the constraint equation again takes the Heun function form
\begin{equation}
2\gamma k(1+k^{2}e^{k\varphi})W''+(3k^{6}\gamma e^{2k\varphi}+k^{2}e^{k\varphi}(2+k^{2}\gamma(\beta+5))+k^{2}\gamma(\beta+2)+2)W'+(3k^{5}e^{2k\varphi}+k^{3}e^{k\varphi}(\beta+5)+k(\beta+2))W=0.
\label{21}
\end{equation}
Now, we rewrite the above equation,
\begin{equation}
x^{2}(1+x^{2})W''+\bigg(3k^{2}\gamma x^{5}+x^{3}(2+k^{2}\gamma(\beta+6))+(k^{2}\gamma(\beta+3)+2)x\bigg)\frac{W'}{\gamma k^{2}}+2\bigg(3x^{4}+x^{2}(\beta+5)+(\beta+2)\bigg)\frac{W}{\gamma k^{2}}=0
\label{22}
\end{equation}
where $k^{2}e^{k\varphi}=x^{2}$. As we can see the obtained Heun equation (\ref{22}) is similar to the Heun equation expressed in (\ref{3}) with the polynomial functions $f_{1}(x)$, $f_{2}(x)$ and $f_{3}(x)$ which are introduced by the relations (\ref{5}) where the coefficients $a_{i}$ (with $i=0,...,17$) are substituted from the equation (\ref{22}). Then, we can find the symmetry operators as,
\begin{equation}
P^{+}=3x^{5}\frac{d}{dx}+\frac{6}{\gamma k^{2}}x^{4},\quad\quad\quad P^{-}=x^{4}\frac{d^{2}}{dx^{2}}+\frac{(2+\gamma k^{2}(\beta+6))}{\gamma k^{2}}x^{3}\frac{d}{dx}+\frac{2(\beta+5)}{\gamma k^{2}}x^{2},\quad\quad\quad P^{0}=\mu x^{4}\frac{d^{2}}{dx^{2}}+\nu x^{3}\frac{d}{dx}+\zeta x^{2}
\label{23}
\end{equation}
where $\mu$, $\nu$ and $\zeta$ take the following forms 
\begin{equation}
\mu=1,\quad\quad\quad \nu=\frac{10+\gamma k^{2}(\beta+16)}{3\gamma k^{2}},\quad\quad\quad \zeta=\frac{8+2\gamma k^{2}(\beta+13)}{3\gamma k^{2}}.   
\label{24}
\end{equation}
In order to make the above generators with the following commutation relations
\begin{equation}
[P^{0},P^{-}]=-P^{+},\quad\quad\quad [P^{0},P^{+}]=18x^{4}P^{-},\quad\quad\quad [P^{+},P^{-}]=-18x^{4}P^{0},
\label{25}
\end{equation}
we need the conditions 
\begin{equation}
2-\gamma k^{2}(\beta+1)=0,\quad\quad\quad -3\gamma^{2}k^{4}(4\beta+7)-8\gamma k^{2}(2\beta-1)+32=0,\quad\quad\quad -2\gamma^{2}k^{4}(8\beta+17)-16\gamma k^{2}(\beta-1)+32=0. 
\label{26}
\end{equation}
As a consequence of the above discussion, we realize that the operators $P^{+}$, $P^{-}$ and $P^{0}$ behave like the operators $K^{+}$ $K^{-}$ and $K^{0}$ of the $sl(1, 1)$ algebra. Therefore, we can conclude the symmetries behind of the non-minimal constant-roll inflation with exponential couplings coupling is in a good agreement with the $sl(1, 1)\sim sp(2, R) \sim so(2, 1)$ algebra which is known as the isomorph structure of $sl(2, R)$. Note that the operator $P^{+}$ is the main responsible for the present generalization of algebra unlike the previous case which all three operators produced same generalizations. 
\section{Conclusion}
As a new approach to describe the inflationary epoch, we have used the constant-roll inflation going beyond the slow-roll approximation by considering a non-negligible $\ddot{\varphi}$ in the Klein-Gordon equation. We have applied the discussed approach for the non-minimal coupling inflationary model which considers a direct interaction between the Ricci scalar and scalar field by adding a non-minimal coupling term in the form of standard inflationary action. To investigate the corresponding model, we have abandoned the conventional method used in constant-roll papers and have employed the first-order formalism coming from supersymmetry concepts. We have described the obtained results in our previous paper addressed in the body of the present manuscript. 

Due to the main goal of this paper, we have tried to find the hidden symmetries of the model through the results obtained in our previous paper \cite{sadeghi}. In order to fulfill this, we have used the symmetry properties of the Heun function instead of the Killing vector approach of the Lie symmetry since the obtained equations of the model have found the form of the Heun differential equation. Hence, first, we have briefly described the Heun function and its symmetry features used in this work. Then, we have studied the considered inflationary model and have engaged the algebra method for the model. As the consequence, we have found that the hidden symmetries of the non-minimal constant-roll inflation in the cases of power-law and exponential couplings are characterized as a generalized form of $sl(2, R)$ and $su(1, 1)$ algebra, respectively.   
\bibliographystyle{ieeetr}
\bibliography{biblo}
\end{document}